\documentstyle[12pt]{article}
\def\ai{\'{\i}}

\def\e3o{e^{3\Omega}}

\def\-3o{e^{-3\Omega}}

\def\be{\begin{equation}}
\def\ee{\end{equation}}

\begin{document}

\baselineskip.33in

\centerline{\large{\bf COMMENTS ON QUANTUM COSMOLOGY}} \centerline{\large 
{\bf WITH EXTRINSIC TIME}}

\centerline{\large{\bf }}

\vskip0.8cm

\centerline{GAST\'ON GIRIBET}

\centerline{\it  Instituto de Astronom\ai a y F\ai sica del Espacio} 
\centerline{\it C.C. 67, Sucursal 28 - 1428 Buenos Aires, Argentina} 
\centerline{\it E-mail: gaston@iafe.uba.ar }

\bigskip

\centerline{CLAUDIO SIMEONE}

\centerline{\it Departamento de F\'{\i }sica, Comisi\'{o}n Nacional de
Energ\'{\i }a At\'{o}mica} \centerline{\it Av. del Libertador 8250 - 1429
Buenos Aires, Argentina} \centerline{\it and} \centerline{\it Departamento
de F\'{\i }sica, Facultad de Ciencias Exactas y Naturales} \centerline{\it 
Universidad de Buenos Aires, Ciudad Universitaria} \centerline{\it 
Pabell\'{o}n I - 1428, Buenos Aires, Argentina} \centerline{\it E-mail:
simeone@tandar.cnea.gov.ar}

\vskip1cm

\noindent

ABSTRACT

\vskip1cm

The {\it closed} de Sitter universe is used to present a way to deal with
the deparametrization and quantization of cosmological models with extrinsic
time.

\vskip1cm


\newpage

\vskip1cm

\section{Introduction}

\bigskip

\noindent An essential property of gravitational dynamics is that the
canonical Hamiltonian vanishes on the physical trajectories of the system,
which then includes a constraint ${\cal H}\approx 0$. This reflects the fact
that the evolution of the gravitational field is given in terms of a
parameter $\tau $ which does not have physical significance. This feature
leads to a fundamental difference between ordinary quantum mechanics and the
quantization of gravitation, because the existence of a unitary quantum
theory is related to the possibility of defining the time as an absolute
parameter. The identification of a global phase time \cite{rafa1} can
therefore be considered as the previous step before quantization \cite{haj}.

For minisuperspace models we have an action functional of the form 
\begin{equation}
S[q^{i},p_{i},N]=\int_{\tau _{1}}^{\tau _{2}}\left( p_{i}{\frac{dq^{i}}{
d\tau }}-N{\cal H}\right) d\tau 
\end{equation}
where $N$ is a Lagrange multiplier enforcing the quadratic Hamiltonian
constraint 
\begin{equation}
{\cal H}=G^{ij}p_{i}p_{j}+V(q)\approx 0,
\end{equation}
with $G^{ij}$ the reduced version of the DeWitt supermetric. The extremal
condition $\delta S=0$ gives the canonical equations 
\begin{equation}
{\frac{dq^{i}}{d\tau }}=N[q^{i},{\cal H}],\ \ \ \ \ \ \ \ \ {\frac{dp_{i}}{
d\tau }}=N[p_{i},{\cal H}].
\end{equation}
The solution of these equations describes the evolution of a spacelike
hypersurface along the timelike direction; the presence the of the
multiplier $N$ introduces an arbitrariness in the evolution which is
associated to a multiplicity of times. From a different point of view, the
constraint ${\cal H}\approx 0$ acts as a generator of gauge transformations
which can be written 
\begin{equation}
\delta _{\epsilon }q^{i}=\epsilon (\tau )[q^{i},{\cal H}],\ \ \ \ \ \delta
_{\epsilon }p_{i}=\epsilon (\tau )[p_{i},{\cal H}],\ \ \ \ \ \delta
_{\epsilon }N={\frac{\partial \epsilon (\tau )}{\partial \tau }}.
\end{equation}

From (3) and (4) we see that the dynamical evolution can be reproduced by a
gauge transformation progressing with time, that is, any two succesive
points on each classical trajectory are connected by a gauge transformation 
\cite{bar}. Hence, the gauge fixation can be thought not only as a way to
select one path from each class of equivalent paths in phase space, but also
as a reduction procedure identifying a time for the system.

Admissible gauge conditions are those which can be reached from any path in
phase space by means of gauge transformations leaving the action unchanged.
Under a gauge transformation defined by the parameters $\epsilon ^{m}$ the
action of a system with constaints $C_{m}$ changes by 
\begin{equation}
\delta _{\epsilon }S=\left[ \epsilon ^{m}(\tau )\left( p_{i}{\frac{\partial
C_{m}}{\partial p_{i}}}-C_{m}\right) \right] _{\tau _{1}}^{\tau _{2}}.
\end{equation}
For ordinary gauge systems, which include constraints that are linear and
homogeneous in the momenta, it is $\delta _{\epsilon }S=0$, and gauge
conditions of the form $\chi (q,p,\tau )=0$ ({\it canonical gauges}) are
admissible. In the case of gravitation, instead, the Hamiltonian constraint
is quadratic in the momenta, and we would have $\delta _{\epsilon }S\neq 0$
unless $\epsilon (\tau _{1})=\epsilon (\tau _{2})=0$ \cite{tei}; then gauge
conditions involving derivatives of Lagrange multipliers as, for example, $
\chi \equiv dN/d\tau =0$ ({\it derivative gauges}) should be used \cite{hall}
. These gauges cannot define a time in terms of the canonical variables. At
the quantum level this has the consequence that the usual Fadeev--Popov path
integral procedure for quantizing gauge systems could not be applied \cite
{htv}.

However, in recent papers \cite{rafa2}-\cite{clo2} we have shown that if the
system results separable the action can be provided with gauge invariance at
the end points by means of a canonical transformation $(q^{i},p_{i})\to
(Q^{i},P_{i})$ which matches the constraint ${\cal H}$ with one of the new
momenta. Then canonical gauges are admissible and a global phase time can be
identified by imposing $\tau -$dependent gauge conditions, and,
simultaneously, the transition amplitude can be obtained by means of the
usual path integral procedure for gauge systems. Moreover, we have shown
that a transition amplitude between two given quantum configurations in
terms of only the original coordinates can be found only in the case that an 
{\it intrinsic time} exists.

A function $t(q^{i},p_{i})$ is a global phase time if $[t,{\cal H}]>0.$
Because the supermetric $G^{ik}$ does not depend on the momenta, a function $
t(q)$ is a global phase time if the bracket 
\begin{eqnarray*}
\lbrack t(q),{\cal H}] &=&[t(q),G^{ik}p_{i}p_{k}] \\
&=&2{\frac{\partial t}{\partial q^{i}}}G^{ik}p_{k}
\end{eqnarray*}
is positive definite. If the supermetric has a diagonal form and one of the
momenta vanishes at a given point of phase space, then no function of its
conjugate coordinate only can be a global phase time. This is something to
be remarked, as in some early works \cite{ryan} the scale factor was used as
time parameter in the obtention of a Schr\"{o}dinger equation for the
minisuperspaces, and this is not licit in general; examples of this are the
cases of Kantowski--Sachs anisotropic universe \cite{clo2} and isotropic
models yielding from the low energy limit of closed bosonic string theory 
\cite{nos}. For a constraint whose potential can be zero for finite values
of the coordinates, the momenta $p_{k}$ can be all equal to zero at a given
point, and $[t(q),{\cal H}]$ can vanish. Hence an intrinsic global phase
time $t(q)$ can be identified only if the potential in the constraint has a
definite sign. In the most general case a global phase time should be a
function including the canonical momenta; in this case it is said that the
system has an extrinsic time $t(q,p)$ \cite{kuch}, because the momenta are
related to the extrinsic curvature. When only an extrinsic time exists we shall have to revise some points of the path integral quantization to which
we are used: as we shall see below, a quantum description in terms of only
the original coordinates may be impossible if we want to work in a theory
with a clear notion of time.

\vskip1cm

\bigskip


\section{A simple model: The {\em closed} de Sitter universe.}

In order to discuss this topic we shall use a simple model which presents
all the peculiarities of the models of physical interest; this model has the
special feature that the transition probability is trivially known since it
has one constraint and only one degree of freedom, so that there is only one
physical state for the system. Despite its simplicity, it is a good example
to understand the quantization with extrinsic time.

Consider the Hamiltonian constraint of the most general empty homogeneous
and isotropic cosmological model: 
\begin{equation}
{\cal H}=-{\frac{1}{4}}e^{-3\Omega }\pi _{\Omega }^{2}-ke^{\Omega }+\Lambda
e^{3\Omega }\approx 0.
\end{equation}
This Hamiltonian corresponds to a universe with arbitrary curvature $k=-1,0,1
$ and non zero cosmological constant; we shall suppose $\Lambda >0$. In the
case $k=0$ we obtain the de Sitter universe; although the absence of matter
makes this universe basically a toy model, it has received considerable
attention because it reproduces the behaviour of models with matter or with
non zero curvature when the scale factor $a\sim e^{\Omega }$ is great
enough. The classical evolution is easy to obtain, and corresponds to an
exponential expansion. In fact, for both $k=0$ and $k=-1$ the potential is
never zero, and then $\pi _{\Omega }$ cannot change its sign. Instead, for
the closed model $\pi _{\Omega }=0$ is possible.

It is convenient to work with the rescaled Hamiltonian $H=e^{-\Omega}{\cal H}
:$ 
\begin{equation}
H=-{\frac{1}{4}}e^{-4\Omega}\pi_\Omega^2 -k+\Lambda e^{2\Omega}\approx 0.
\end{equation}
The constraints $H$ and ${\cal H}$ are equivalent because they differ only
in a positive definite factor.

We shall turn the parametrized system asociated to the Hamiltonian (7) into
an ordinary gauge system by means of two succesive canonical
transformations. The $\tau -$independent Hamilton--Jacobi equation for the
Hamiltonian $H$ is 
\begin{equation}
-\left( {\frac{\partial W}{\partial \Omega }}\right) ^{2}-4ke^{4\Omega
}+4\Lambda e^{6\Omega }=4e^{4\Omega }E.
\end{equation}
Matching $E=\overline{P}_{0}$ we obtain the solution 
\begin{equation}
W(\Omega ,\overline{P}_{0})=2sign(\pi _{\Omega })\int d\Omega e^{2\Omega }%
\sqrt{\Lambda e^{2\Omega }-k-\overline{P}_{0}},
\end{equation}
which is the generating function of the canonical transformation $(\Omega
,\pi _{\Omega })\to (\overline{Q}^{0},\overline{P}_{0})$ defined by 
\begin{equation}
\overline{Q}^{0}={\frac{\partial W}{\partial \overline{P}_{0}}}=-sign(\pi
_{\Omega })\Lambda ^{-1}\sqrt{\Lambda e^{2\Omega }-k-\overline{P}_{0}}, \ \ \ \ \ \ \ \ \ \overline{P}_{0}=H.
\end{equation}
Then we define the function $F=\overline{Q}^{0}P_{0}+f(\tau )$ which
generates the second canonical transformation yielding a non vanishing true
Hamiltonian $h=\partial f/\partial \tau $ and $Q^{0}=\overline{Q}^{0}$, $
\overline{P}_{0}=P_{0}.$ This second transformation may seem to be an
unnecesary sofistication, but it plays a central role in the case of
cosmological models with true degrees of freedom, as it allows the new
coordinates $Q$ associated to observables to be fixed at arbitrary values at
the boundaries. If possible, $f$ should be chosen in such a way that the
reduced Hamiltonian is conservative \cite{clo3}.

The variables $Q^{0}$ and $P_{0}$ describe the gauge system into which the
model has been turned. Therefore the gauge can be fixed by means of a $\tau -
$dependent canonical condition like $\chi \equiv Q^{0}-T(\tau )=0$ with $T$
a monotonic function of $\tau $. Then as $[Q^{0},P_{0}]=1$ we can define the
global phase time as 
\begin{equation}
t=Q^{0}|_{P_{0}=0}=-sign(\pi _{\Omega })\Lambda ^{-1}\sqrt{\Lambda
e^{2\Omega }-k}
\end{equation}
As on the constraint surface $P_{0}=0$ we have 
\begin{equation}
\pi _{\Omega }= 2 sign(\pi _{\Omega })e^{2\Omega }\sqrt{\Lambda e^{2\Omega }-k}
,
\end{equation}
(so that in the case $k=1$ the natural size of the configuration space is
given by $\Omega \geq -\ln (\sqrt{\Lambda })$ \cite{haj}) we can write 
\begin{equation}
t(\Omega ,\pi _{\Omega })=-{1\over 2}\Lambda ^{-1}e^{-2\Omega }\pi _{\Omega },
\label{time}
\end{equation}
which is in agreement with the time obtained by matching the model with the
ideal clock \cite{rafa3,hernan}. It is interesting to notice that
this expresion for the time can be also obtained by demanding that its
functional form presents no ambiguities under different ways of factorizing
the Hamiltonian constraint. 

Now an important difference between the cases $
k=-1$ and $k=1$ arises: for $k=-1$ the potential has a definite sign, and
the constraint surface splits into two disjoint sheets given by (12). In
this case the evolution can be parametrized by a function of the coordinate $
\Omega $ only, the choice given by the sheet on which the system remains: if
the system is on the sheet $\pi _{\Omega }>0$ the time is $t=-\Lambda ^{-1}
\sqrt{\Lambda e^{2\Omega }+1}$, and if it is on the sheet $\pi _{\Omega }<0$
we have $t=\Lambda ^{-1}\sqrt{\Lambda e^{2\Omega }+1}$. The
deparametrization of the flat model is completely analogous. For the closed
model, instead, the potential can be zero and the topology of the constraint
surface is no more equivalent to that of two disjoint planes. Although for $
\Omega =-\ln (\sqrt{\Lambda })$ we have $V(\Omega )=0$ and $\pi _{\Omega }=0,
$ it is easy to verify that $d\pi _{\Omega }/d\tau \neq 0$ at this point.
Hence, in this case the coordinate $\Omega $ does not suffice to parametrize
the evolution, because the system can go from $(\Omega ,\pi _{\Omega })$ to $
(\Omega ,-\pi _{\Omega })$; therefore we must necessarily define a global
phase time as a function of both the coordinate and the momentum.

The system has one degree of freedom and one constraint, so that it is pure
gauge. In other words, there is only one physical state, in the sense that
from a given point in the phase space we can reach any other point on the
constraint surface by means of a finite gauge transformation. For this
reason, if the deparametrization procedure is consistent it should be
possible to verify that the transition probability written in terms of the
variables which include a globally well defined time is equal to unity

The quantization proceeds as in references \cite{rafa2}-\cite{clo2}, and the
observation above is reflected in the fact that we obtain 

\begin{eqnarray}
<Q_{2}^{0},\tau _{2}|Q_{1}^{0},\tau _{1}> &=&\int DQ^{0}DP_{0}DN\delta
(Q^{0}-\tau )\exp \left( i{\int_{\tau _{1}}^{\tau _{2}}\left[ P_{0}{\frac{%
dQ^{0}}{d\tau }}-NP_{0}-{\frac{\partial f}{\partial \tau }}\right] d\tau }%
\right)  \nonumber \\
&=&\int DQ^{0}DP_{0}\delta (P_{0})\delta (Q^{0}-\tau )\exp \left(
i\int_{\tau _{1}}^{\tau _{2}}\left[ P_{0}{\frac{dQ^{0}}{d\tau }}-{\frac{%
\partial f}{\partial \tau }}\right] d\tau \right)  \nonumber \\
&=&\exp \left( -i\int_{\tau _{1}}^{\tau _{2}}{\frac{\partial f}{\partial
\tau }}d\tau \right) ,
\end{eqnarray}
and then the probability for the transition from $Q_{1}^{0}$ at $\tau _{1}$
to $Q_{2}^{0}$ at $\tau _{2}$ is 
\begin{equation}
\left| <Q_{2}^{0},\tau _{2}|Q_{1}^{0},\tau _{1}>\right| ^{2}=1.
\end{equation}
When the model is open or flat the coordinates $\Omega $ and $Q^{0}$ are
uniquely related, and the result can be easily understood in the sense that
once a gauge is fixed there is only one possible value of the scale factor $
a\sim e^{\Omega}$ at each $\tau $. But in the case of the closed model we
have seen that this is not true: at each $\tau $ there are two possible
values of $\Omega $; instead, there is only one possible value of $\pi
_{\Omega }$ at each $\tau $. Hence the transition probability in terms of $
Q^{0}$ does not correspond to the evolution of the coordinate $\Omega $, but
rather of its derivative; then we conclude that the amplitude $
<Q_{2}^{0},\tau _{2}|Q_{1}^{0},\tau _{1}>$ corresponds to an amplitude $<\pi
_{\Omega ,2}|\pi _{\Omega ,1}>$, and we have 
\begin{equation}
|<\pi _{\Omega ,2}|\pi _{\Omega ,1}>|^{2}=1.
\end{equation}
The fact that $<Q_{2}^{0},\tau _{2}|Q_{1}^{0},\tau _{1}>$ is not equivalent
to $<\Omega _{2}|\Omega _{1}>$ is natural, as the nonexistence of an
intrinsic time makes impossible to find a globally good gauge condition
giving $\tau $ as a function of $\Omega $ only \cite{clo3}. But precisely
for this reason, this should not be taken as a failure of the quantization
procedure, because a characterization of the states in terms of only the
original coordinates is not correct if we want to retain a clear notion of
time on the whole evolution.

\section{Discussion}

It is common to regard an intrinsic time as more {\it natural}, and the
necessity of defining an extrinsic time as a problematic peculiarity.
However, this is perhaps only a consequence of working with simple
parametrized systems like, for example, the relativistic particle; the
formalism for these systems, when put in a manifestly covariant form, has
the time included among the coordinates, and the evolution is given in terms
of a physically meaningless time parameter. But while for these systems the
time coordinate always refers to an external clock, this is clearly not the
case in cosmology; for example, in the case of pure gravitational dynamics
the coordinates are the elements of the metric $g_{ab}$ over spatial slices,
and in principle there is not necessarily a connection between $g_{ab}$ and
anything {\it external}. Rather, such a relation can be thought to exist for
the derivatives $dg_{ab}/d\tau $ of the metric, as they appear in the    
expression for the extrinsic curvature $K_{ab}$ which describes the
evolution of spacelike 3-dimensional hypersurfaces in 4-dimensional
spacetime. If no matter fields are present the canonical momenta are given
by 
\begin{equation}
\pi _{i}\equiv p^{ab}=-2G^{abcd}K_{cd},
\end{equation}
and then one must expect the momenta to appear in the definition of a global
phase time. The existence of a time in terms of only the coordinates should
therefore be understood as a sort of an {\it accident} related to the fact
that, in some special cases which do not represent the general features of
gravitation, there exists a relation that enables to obtain the coordinates
in terms of the momenta with no ambiguities.

This means that we must revise some points of the path integral quantization
to which we are used. Indeed, the system considered in this brief note
represents an example of a cosmological model for which the characterization
of the quantum states must include the momenta.

It is worth noting that at the quantum level the definition of an extrinsic
time in terms of the functional form (\ref{time}) is not sufficient; in
fact, it is also necessary to propose a prescription for the operatorial
order between coordinates and momenta to give a precise definition of time.
Indeed, it is possible to verify that an ordering which leads to define an
extrinsic time for the closed de Sitter universe is given by 
\begin{equation}
\hat{t}\sim \hat{\pi}_{\Omega }e^{-2\hat{\Omega}}.
\end{equation}
We must remark that different orderings generate in the conmutator $[\hat{t},
\hat{H}]$ linear terms in $\pi _{\Omega }$, which lead to a non definite
sign.

At this point, it could be interesting to comment that the description of
the evolution of cosmological models in terms of the momenta is not a
particular property of quantization; in fact, this is a common feature of
classical cosmology where it is usual to deal with cosmological dynamics
where the time is given in terms of the Hubble constant H, namely 
\begin{equation}
t_{{\rm H}}={\rm H}^{-1}\sim -{\frac{e^{3\Omega }}{\pi _{\Omega }}}.
\label{class}
\end{equation}
Note that on the constraint surface the time (\ref{time}) can be
rewritten as  
\begin{equation}
t\sim \frac{ke^{2\Omega }}{\pi _{\Omega }}-\frac{\Lambda e^{4\Omega }}{\pi
_{\Omega }},
\end{equation}
which has an analogous form.

The problem of time in quantum gravity is an important open question. In
this paper we have identified time variables for a particular minisuperspace
model which is of interest in quantum cosmology by means of a systematic
procedure to deparametrize the constrained system, and we have given a
consistent quantization with a clear notion of time. We have pointed out the
relevance of the study of the structure of the phase space in terms of the
extrinsic curvature within the context of the procedure to find a global
phase time for deparametrizing cosmological models which do not admit an
intrinsic time.

\vskip0.8cm

{\bf Acknowledgement}

\medskip

This work was supported by CNEA and CONICET.

\end{document}